\documentclass[journal]{IEEEtran}

\usepackage{cite}
\usepackage{longtable}
\usepackage{amsmath,amssymb,amsfonts}
\usepackage{textcomp}
\def\BibTeX{{\rm B\kern-.05em{\sc i\kern-.025em b}\kern-.08em
    T\kern-.1667em\lower.7ex\hbox{E}\kern-.125emX}}
\usepackage{booktabs}
\usepackage{multirow}
\usepackage{orcidlink}
\usepackage{hyperref}
\hypersetup{hidelinks}

\title{Non-invasive Techniques for Flow Rate Measurement in Water Pipes: Protocol for a Systematic Review}
\author{Juan Diego Belesaca $^{1, 2}$ \orcidlink{0000-0001-8609-0358},
        Fabian Astudillo~Salinas $^2$ \orcidlink{0000-0001-7644-0270}\\
        $^{1}$ \textit{Computer Science Department}, Universidad de Cuenca, Cuenca, Ecuador\\
        $^{2}$ \textit{Department of Electrical Engineering, Electronic and Telecommunications}, Universidad de Cuenca, Cuenca, Ecuador\\
        E-mails: \{juan.belesca, fabian.astudillos\} @ucuenca.edu.ec

\vspace{-0.9cm}
        
\thanks{This work was supported by the Universidad de Cuenca through the Vice-Rectorate for Research and Innovation under the project “Prediction of Water Consumption in Households Using Non-Invasive Vibration Sensors and Predictive Algorithms Adapted to Socioeconomic Contexts.” The author also acknowledges the financial support of the Universidad de Cuenca through the postgraduate scholarship of the Doctorate in Applied Computer Science.}}
\begin{document}
\maketitle

\begin{abstract}
Accurate, non-invasive flow measurement is imperative for efficient water resource management and leak detection in distribution systems. Despite the advent of diverse external sensing technologies, a paucity of consolidated evidence exists regarding their comparative performance, energy efficiency, and applicability in varied operational contexts. The document delineates the protocol for a systematic literature review (SLR) that aims to identify, evaluate, and synthesize the extant evidence on non-invasive flow monitoring techniques for piped networks. Adhering to the Kitchenham methodology, the review will investigate the accuracy, precision, and energy consumption of prevailing solutions, such as ultrasonic and accelerometer-based systems. The analysis will also assess the impact of signal processing and machine learning (ML) algorithms on enhancing system capabilities. The objective of this study is to map the state-of-the-art, identify key research gaps, and provide an empirical foundation to direct future research toward operational deployment. 
\end{abstract}


\section*{Review Highlights}
\begin{itemize}
    \item A systematic literature review was conducted to synthesize existing evidence on non-invasive flow measurement technologies.
    \item The review methodically extracted and analyzed data on the performance, efficiency, and applicability of current methods.
    \item A comprehensive evidence map was meticulously constructed to provide a comprehensive overview of the current state of technology, highlighting critical gaps that necessitate further investigation and subsequent research and deployment.

\end{itemize}

\section{Introduction}
The sustainable management of water resources has become an urgent global challenge, particularly in the context of climate change, rapid urbanization, and population growth. The increasing water demand has been shown to exert pressure on existing supply systems. This phenomenon is further compounded by the unpredictability of climate-related events, which serve to further complicate the balance between availability and consumption. In this scenario, innovative solutions are required to optimize residential water use and to provide decision-making tools for both households and policymakers.

This systematic review is motivated by the need to evaluate the current state of non-invasive technologies for water flow measurement in domestic environments. The objective of this study is to synthesize extant evidence and identify technological advances that can enable accurate monitoring and prediction of residential water consumption without altering current infrastructure. These innovations are imperative for the promotion of behavioral modifications, the reduction of waste, and the support of long-term water sustainability strategies.

Recent research has explored non-invasive sensor technologies as a practical and cost-effective alternative to traditional flow meters. For instance, studies have demonstrated the use of vibration and temperature sensors to estimate hot water consumption in households \cite{1}. In a similar vein, Venkata and Navada \cite{2} employed a combination of vibration sensing and Fourier analysis, correlating frequency magnitudes with flow rate. They enhanced the accuracy of this approach using neural network models. Other works have advanced the integration of signal processing to estimate water flow \cite{3}. However, many of these systems remain invasive or are only applicable in controlled environments, limiting their real-world applicability.

In addition to vibration-based approaches, alternative solutions have been investigated. Optical sensors, for instance, have been utilized in the measurement of multiphase flow \cite{4}, while vibration sensors have been employed to detect particulate events in water–gas mixtures \cite{5}. Furthermore, high-resolution imaging has been integrated with vibration detection to enhance the accuracy of the system \cite{6}. In industrial applications, the implementation of optical and fiber Bragg grating sensors has been observed to facilitate the measurement of turbulence and complex flow conditions \cite{7,8}. Even though these technologies are designed for large-scale or high-precision environments, they underscore the versatility and potential of non-invasive sensing methods.

Notwithstanding these advancements, challenges persist regarding cost, scalability, and precision in domestic contexts. The household environment necessitates solutions that can concurrently monitor multiple consumption points, ensure precision while maintain cost-effectiveness and ease of implementation. Moreover, water management strategies must address not only immediate demand but also broader ecological and sustainability objectives \cite{9,10}.

The objective of this systematic review is to provide a comprehensive overview of the methodologies, sensor technologies, and data-driven approaches applied to non-invasive residential water monitoring. The objective of the review is twofold: first, to identify research gaps, particularly in the integration of machine learning algorithms with sensor networks to predict water consumption under diverse socio-economic conditions; and second, to provide a comprehensive overview of the state of the art in this field. The knowledge consolidated here will contribute to the development of innovative, sustainable solutions with direct implications for urban and rural water resource planning, public policy, and the preservation of ecosystems.

\section{Method Details}

The review employs a systematic mapping methodology based on the three-stage process proposed by Kitchenham and Charters: Planning, Conducting, and Reporting \cite{11}. The following sections detail the protocol designed for this review.

\subsection{Planning stage}

The planning stage consists of six distinct steps: formulating research questions, defining a search strategy, establishing criteria for primary study selection, creating a quality assessment protocol, designing a data extraction strategy, and selecting synthesis methods.

The research is guided by a primary research question (RQ) and four supplementary sub-questions (RSQs). To structure these questions, the Population, Intervention, Comparison, and Outcome (PICO) framework \cite{12} was employed. The primary research question (RQ) is: What non-invasive devices and techniques can be used for monitoring water flow and consumption, and how do they compare in terms of performance? 

The PICO elements for the main RQ are defined as follows:

\begin{itemize}
    \item Population: water flow and consumption monitoring.
    \item Intervention: non-invasive devices and techniques.
    \item Comparison: other non-invasive techniques (implicit, among the non-invasive techniques found).
    \item Outcome: performance, including measurement accuracy, device robustness, ease of implementation, adaptability to different pipe types and dimensions, device cost, and energy consumption.
\end{itemize}

To provide a granular analysis, the RQ is decomposed into four specific research sub-questions (RSQs):

RSQ1 – Types of sensors: What types of sensors are used in non-invasive monitoring of water flow and consumption, and what are their key characteristics? This question investigates non-invasive sensors (Intervention) within the context of water monitoring (Population) to identify their characteristics, such as accuracy and cost (Outcome).

RSQ2 – Application environments: In which application environments are non-invasive water monitoring systems deployed (e.g., residential, distribution networks, industrial settings), and how does the context influence system design and performance? This question examines the deployment of non-invasive systems (Intervention) across different environments (Population) to understand the impact on their design and performance (Outcome).

RSQ3 – Signal processing and communication techniques: What signal processing and communication techniques are applied to data from non-invasive water monitoring sensors, and how do they contribute to system performance and reliability? This question focuses on signal processing and communication techniques (Intervention) applied to sensor data (Population) to evaluate their effect on system performance and reliability (Outcome).

RSQ4 – Use of Machine Learning (ML): How is machine learning integrated into non-invasive water monitoring systems, and what roles does it play (e.g., prediction, adaptation, classification, or improving accuracy)? This question explores the use of machine learning techniques (Intervention) in non-invasive systems (Population) to identify their applications and benefits (Outcome).

\subsubsection{Search Strategy}

The search for primary studies is conducted across several prominent digital libraries, including the ACM Digital Library, IEEE Xplore, Springer Nature Link, and ScienceDirect. In addition to these automated searches, a manual search is performed in targeted core conferences and high-impact journals to ensure comprehensive coverage of the relevant literature. In addition to these automated searches, manual searches are performed in targeted core conferences and high-impact journals to ensure comprehensive coverage of the relevant literature. 
To complement these searches and ensure comprehensive coverage of the literature, the search strategy is augmented with a snowballing technique. This procedure involves systematically reviewing the reference lists of all primary studies that meet the inclusion criteria.

Any cited work that is potentially relevant to the research questions is then evaluated using the same inclusion and exclusion criteria defined in this protocol. To maintain a focused and manageable scope, this process is restricted to a single iteration. Consequently, if a paper identified via snowballing is a literature review, its bibliography is not examined to initiate a second layer of the search. Table 1 and Table 2 detail the specific venues selected for manual review using the snowball technique.

\begin{table}
    \centering
    \caption{Core Conferences for Manual Search}
    \begin{tabular}{p{4cm}cc}
        \toprule
        \centering{\textbf{Title}}& \textbf{Acronym} & \textbf{SJR (CORE)}\\
        \midrule
        Conference Record of the Asilomar Conference on Signals, Systems and Computers & ACSSC & 0.337 (USA)\\
        IEEE International Instrumentation and Measurement Technology Conference  & I2MTC & 0.258\\
        ACM SIGKDD International Conference on Knowledge Discovery and Data Mining & ACM SIGKDD & 0,692(A*)\\
        Fluid-Structure-Sound Interactions and Control & FSSIC & -\\
        \bottomrule
    \end{tabular}
    \label{tab:1}
\end{table}

\begin{table}
    \centering
    \caption{Journals Selected for Manual Search}
    \begin{tabular}{p{4.5cm}cc}
        \toprule
        \centering{\textbf{Title}}& \textbf{Country} & \textbf{Ranking}\\
        \midrule
        IEEE Sensors Journal & United States & Q1\\
        IEEE Transactions on Ultrasonics, Ferroelectrics, and Frequency Control  & United States & Q1\\
        Water Resources Management & Netherlands & Q1\\
        Advanced Engineering Informatics & United Kingdom & Q1\\
        Environmental Modelling \& Software & United Kingdom & Q1\\
        Experimental Thermal and Fluid Science & United States & Q1\\
        \bottomrule
    \end{tabular}
    \label{tab:2}
\end{table}

A structured search string was developed by combining concepts with Boolean operators, as outlined in Table 3.

\begin{table}
    \centering
    \caption{Search String Construction}
    \begin{tabular}{p{2cm}p{2cm}cp{2cm}}
        \toprule
        \textbf{Concept} & \textbf{Sub-string} & \textbf{Connector} & \textbf{Alternative}\\
        \midrule
        Non-invasive & Non-invasive & AND & Non-intrusive\\
        Flow measurement & Flow measurement & AND & Flow rate\\
        Water & Water & AND & Fluid\\
        Medical applications & Medic* & NOT & Clinical, blood, arter*\\
        Multiphase fluids & Multiphase & NOT & *phase, gas, oil, plant*\\
        \bottomrule
    \end{tabular}
    \label{tab:placeholder}
\end{table}

The final search string applied is: \textit{(non-invasive OR non-intrusive) AND (flow measurement OR flow rate) AND (water OR fluid) NOT (blood OR arter* OR medic* OR clinical) NOT (multiphase OR *phase OR gas OR oil OR air OR plant*)} 

The search string is applied to the metadata (title, abstract, and keywords) of articles in each digital library, with syntax adapted as required by each platform. The search covers publications from 2010 to 2025. This period was selected to capture the evolution of non-invasive technologies following key regulatory and standardization milestones, such as the EU Water Framework Directive (2000/60/EC) \cite{13} and ISO 4064:2005 \cite{14}, which established foundational requirements for water monitoring and measurement accuracy. The timeframe allows for an analysis of the technological transition from early prototypes to mature solutions, while focusing on the impact of these foundational regulations. 

\subsubsection{Primary Study Selection}

the selection of primary studies is a multi-author process designed to ensure objectivity. Three researchers independently evaluate each retrieved study based on its title, abstract, and keywords. Any discrepancies in selection are resolved through discussion and consensus after examining the full text of the paper in question.

A study is included if it meets at least one of the following criteria:
\begin{itemize}
    \item Studies that present information on non-invasive techniques for flow rate measurement
    \item Studies that present non-invasive sensors for flow rate measurement in water pipes
    \item Studies that presenting research in non-invasive techniques for detection of events such as leaks in water pipes
\end{itemize}

A study is excluded if it meets at least one of the following criteria:
\begin{itemize}
    \item Studies that present non-invasive techniques for flow rate measurement applied in transparent transport pipes
    \item Introductory papers for short papers, books and workshops.
    \item Duplicate reports of the same study in different sources.
    \item Short papers with less than five pages.
\end{itemize}

\subsubsection{Quality Assessment}

a quality assessment is performed on all included studies to evaluate their rigor and relevance. This assessment utilizes a questionnaire based on a three-point Likert scale, which is composed of two subjective and two objective questions.

The subjective questions assess whether the study discusses non-invasive techniques and sensor networks for flow measurement. Responses are scored as "I agree" (10 points), "Partially" (5 points), or "I do not agree" (0 points).

The first objective question evaluates the relevance of the publication venue. A score of Very Relevant (10 points) is assigned to papers rated A* or A in the CORE classification, published in JCR-indexed journals, or identified during the manual search. A score of Relevant (5 points) is given to papers rated B or C by CORE, published in non-JCR journals, or are theses and technical reports. A score of Not so Relevant (0 points) is assigned to papers from unindexed conferences.

The second objective question assesses the study's impact based on citations. For papers published before 2017, a high rating (10 points) is given for more than five citations, a medium rating (5 points) for 1–5 citations, and a low rating (0 points) for no citations. For papers published in or after 2017, a rating of potentially high (10 points) is given if the paper has been cited, and potentially medium (5 points) if it has not yet been cited.

\subsubsection{Data Extraction Strategy}

a structured data extraction strategy is employed to systematically gather information relevant to each research sub-question. This uniform approach ensures consistency in data collection across all selected studies and facilitates subsequent classification and analysis. The specific data points extracted for each RSQ are detailed in Table 4.

\begin{table*}[htbp]
  \centering
  \caption{Data Extraction Form}
    \begin{tabular}{cp{20em}p{30em}}
    \toprule
    \multicolumn{3}{p{60em}}{\textbf{SRQ 1: What types of sensors are used in non-invasive monitoring of residential water flow and consumption, and what are their key characteristics?}} \\
    \midrule
            &           & Accelerometer \\
            &           & Acoustic receptor \\
            &           & Electromagnetic \\
    EC1     &  Sensor   & Optic fiber \\
            &           & Pressure \\
            &           & Ultrasonic \\
            &           & Others \\
    \midrule
            &           & Prototype \\
    EC2     & Solution offered & End device \\
            &           & Simulation \\
            &           & Others \\
    \midrule
            &           & High precision (Error $ \leq \pm 0.5\%$) \\
    EC3     & Precision& Mean precision (Error ±0.5\% to ±2\%) \\
            &           & Low precision (Error ±2\% to ±5\%) \\
            &           & Very low precision (Error $> ±5\%$) \\
    \midrule
            &           & Galvanized steel \\
    EC4     & Type of piping applied     & PVC \\
            &           & Coper \\
            &           & Other \\
    \midrule
            &           & Ultra-low consumption (µW to mW) \\
    EC5     &  Energy Consumption     & Low consumption (10 mW to 1 W) \\
            &           & Moderate consumption (1 W to 10 W) \\
            &           & High consumption (>10 W) \\
    \midrule
    \multicolumn{3}{p{60em}}{\textbf{SRQ 2: In which application environments are non-invasive water monitoring systems deployed (e.g., residential, distribution networks, industrial settings), and how does the context influence system design and performance?}} \\
    \midrule
            &           & Residential \\
    EC6     & Environments  & Distribution network \\
            &           & Industrial setting \\
            &           & Others \\
    \midrule
    \multicolumn{3}{p{60em}}{\textbf{SRQ 3: What signal processing and communication techniques are applied to data from non-invasive water monitoring sensors, and how do they contribute to system performance and reliability?}} \\
    \midrule
            &           & Filtering \\
            &           & Amplification \\
    EC7     & Signal Processing  & Envelope analysis \\
            &           & Spectral/frequency processing \\
            &           & Feature extraction and analysis \\
            &           & Others \\
    \midrule
            &           & Bluetooth \\
            &           & LoRa \\
    EC8     & Communication  & WIFI \\
            &           & Wired \\
            &           & not specified \\
    \midrule
    \multicolumn{3}{p{60em}}{\textbf{SRQ 4: How is machine learning integrated into non-invasive water monitoring systems, and what roles does it play (e.g., prediction, adaptation, classification, or improving accuracy)?}} \\
    \midrule
            &           & Supervised Learning \\
            &           & Unsupervised Learning\\
    EC9     & Machine learning algorithms & Reinforcement Learning \\
            &           & Not used \\
            &           & Others \\
    \midrule
            &           & Correction of measurements \\
            &           & Prediction \\
    EC10    &   Machine learning role    & Classification \\
            &           & Event detection \\
            &           & Not used \\
            &           & Others \\
    \bottomrule
    \end{tabular}%
  \label{tab:addlabel}%
\end{table*}%

\subsubsection{Data Synthesis Methods}

Both quantitative and qualitative synthesis methods are applied to analyze the extracted data. The quantitative synthesis involves counting the number of primary studies for each category defined in the data extraction form and tracking publication frequency per year and source. The results are visualized using bubble plots to illustrate the frequency of combined findings from different research sub-questions, and a geographical report maps the number of papers by region or country.

The qualitative synthesis focuses on discussing several representative studies for each research sub-question. These studies are selected based on their quality assessment scores to provide a narrative overview of the key findings, challenges, and trends in the research field.

\subsection{Conducting the Review}

In the conducting stage, the review protocol is systematically applied to yield a set of relevant primary studies. The initial search results are filtered according to the defined inclusion and exclusion criteria. During this process, care is taken to identify and consolidate duplicate studies. If a study is published in more than one venue, only the most complete version is retained. If a study appears in multiple digital libraries, it is counted only once according to a predefined source order: ACM, IEEE Xplore, Springer Link, and Science Direct. The outcome of this stage is a final list of selected studies for data extraction and synthesis.

\subsection{Reporting the Results}

The final stage of the methodology is the reporting of the findings. The results of the systematic mapping are structured and communicated in a clear and effective manner to ensure their dissemination to the scientific community and other interested parties.

\section{Method Validation}

The protocol outlined herein was meticulously designed to ensure a rigorous, transparent, and repeatable systematic review of the existing literature. The validity of the systematic review is grounded in its adherence to the established guidelines for such reviews proposed by Kitchenham and Charters \cite{11}. These guidelines are widely accepted within the fields of engineering and computer science.

The robustness of the evidence base is established through a multi-pronged search strategy. The protocol combines a comprehensive automated search across four leading scientific databases with targeted manual searches of high-impact journals and core conferences in the field. This hybrid approach is designed to maximize coverage and mitigate the risk of overlooking key studies that may not be captured by the automated search string alone. Furthermore, the incorporation of a single iteration snowballing technique provides an additional layer of validation, ensuring that the final set of studies is as complete as possible by leveraging the reference lists of already included papers.

In order to minimize the potential for researcher bias and ensure the reliability of the study selection process, this protocol mandates that each potential study be independently evaluated by at least two researchers based on the predefined inclusion and exclusion criteria. Discrepancies are addressed through a consensus discussion process, thereby ensuring that the final selection is the product of meticulous and collective judgment. The subsequent quality assessment stage serves to further bolster the method's validity by means of a systematic evaluation of the rigor and relevance of each included study, thereby enabling a weighted and critical synthesis of the available evidence. The protocol is structured to yield a comprehensive, unbiased, and verifiable overview of the research landscape on non-invasive water monitoring systems by integrating these validation mechanisms directly into the research process.

\section{Limitations}
While this systematic mapping protocol is designed to be comprehensive and rigorous, it is important to acknowledge several potential limitations. The search strategy, while comprehensive, is constrained to four major digital libraries and a curated list of journals and conferences. However, it is important to note that this approach may result in the exclusion of relevant studies published in other databases, non-indexed venues, or presented as "grey literature," such as technical reports, dissertations, or patents. Moreover, the search is constrained to publications written in English. This may introduce a language bias by omitting valuable contributions from non-Anglophone researchers.
The construction of the search string was meticulously developed to maximize relevance; however, it carries an inherent risk of being overly restrictive. The specific keywords and Boolean operators employed to delineate the scope may unintentionally exclude primary studies that utilize alternative terminology to describe relevant technologies. Notably, the NOT clauses, which were designed to eliminate studies from the medical and multiphase fluid domains, may inadvertently omit interdisciplinary research. Such research, while centered on disparate areas, could potentially contain applicable techniques or insights.
The data extraction process is contingent upon a predefined classification schema. While this structured approach is essential for systematic synthesis and quantitative analysis, it may oversimplify the nuances of the primary studies. For instance, the categorization of continuous variables, such as measurement accuracy or energy consumption, into discrete ranges has the potential to obscure the fine-grained details reported by the original authors.
Notwithstanding these limitations, it is anticipated that this protocol will generate a highly representative and reliable overview of the peer-reviewed literature in the designated domain. The multi-faceted search strategy, in combination with a rigorous, multi-researcher selection and quality assessment process, provides a strong foundation for a thorough and unbiased synthesis of the current state-of-the-art in non-invasive water monitoring systems.

\section{Conclusion}
This document presents a thorough and meticulous protocol for a systematic review study on non-invasive technologies for water monitoring. The methodology is grounded in the established guidelines of Kitchenham and Charters, ensuring a transparent, repeatable, and robust process designed to minimize bias and comprehensively cover the relevant literature.

The execution of this protocol is anticipated to yield a comprehensive map of the research landscape, identifying current trends, dominant technologies, and significant research gaps. The resulting synthesis will provide a solid evidence base to guide future research and development in this field. This work is intended to serve as a foundational resource for academics and practitioners seeking to advance the state-of-the-art in efficient and intelligent water resource management.

\hspace{1cm}

\textbf{CRediT author statement:}

\textbf{J. D. Belesaca:} Methodology, Formal analysis, Investigation, Resources, Data curation, Writing - Original Draft. \textbf{F. Astudillo-Salinas:} Conceptualization, Validation, Formal analysis, Writing - Review \& Editing, Supervision, Project administration, Funding acquisition

\section*{Acknowledgment}
This research was supported by the Vice-Rectorate for Research and Innovation at the University of Cuenca. The financial support for this research was provided by the project entitled "Prediction of water consumption in homes using non-invasive vibration sensors and predictive algorithms adapted to socioeconomic contexts."

\ifCLASSOPTIONcaptionsoff
  \newpage
\fi

\bibliographystyle{ieeetr} 
\bibliography{IEEEabrv}

\end{document}